\documentclass[11pt]{article}
\def\be{\begin{equation}}
\def\ee{\end{equation}}
\def\la{\label}
\def\bea{\begin{eqnarray}}
\def\eea{\end{eqnarray}}
\def\non{\nonumber}
\def\ci{\cite}
\def\la{\label}
\def\bib{\bibitem}

\def\lm{\lambda}
\def\Lm{\Lambda}

\def\gm{\gamma}

\def\Omp{\Omega_\phi}
\def\Om{\Omega}
\def\wp{w_\phi}
\def\Ompo{\Omega_{\phi o}}
\def\wpo{w_{\phi o}}
\def\weff{w_{eff}}

\def\fr{\frac}
\def\pp{\partial}

\usepackage{graphicx}

\begin{document}

\begin{flushright}
  astro-ph/0106316\\
  IFUNAM-FT-2001-01\\
\end{flushright}

\vspace{15mm}

\begin{center}
   {\Large \bf  Natural Quintessence with Gauge Coupling Unification}
\end{center}

\vspace*{0.7cm}

\begin{center}
{\bf A. de la Macorra\footnote{e-mail: macorra@fenix.ifisicacu.unam.mx} and C. Stephan-Otto
\footnote{e-mail: stephan@servidor.unam.mx}}
\end{center}

\vspace*{0.1cm}

\begin{center}
\begin{tabular}{c}
{\small Instituto de F\'{\i}sica, UNAM}\\ {\small Apdo. Postal
20-364, 01000  M\'exico D.F., M\'exico}\\
\end{tabular}
\end{center}

\vspace{1 cm}

\begin{center}
{\bf ABSTRACT}
\end{center}
{\small   We show that a positive accelerating universe can be
obtained simply by the dynamics of a non-abelian gauge group. It
is the condensates of the chiral fields that obtain a negative
power potential, below the condensation scale, and allow for  a
quintessence interpretation of these fields. The only free
parameters in this model are $N_c$ and $N_f$ and the number of
dynamically gauge singlet bilinear fields $\phi$ generated below the
condensation scale. We show that it is possible to have
unification of all coupling constants, including   the standard
and non standard model couplings, while having an acceptable
phenomenology of $\phi$ as the cosmological constant. This is done
without any fine tuning of the initial conditions. The problem of
coincidence (why the universe has only recently started an
accelerating period) is not solved but it is put at the same level
as what the particle content of the standard model is.}


\noindent \rule[.1in]{14.5cm}{.002in}

\thispagestyle{empty}

\setcounter{page}{0} \vfill\eject

In the past few years different observations  have lead to conclude
that the universe is flat and  filled with an energy density with
negative pressure, a cosmological constant \ci{SN1a},\ci{CMBR}. The
cosmological constant is perhaps best understood, from an elementary
particle point of view, as the contribution from a scalar field that
interacts with all other fields only gravitationally, i.e.
quintessence \ci{tracker}. Recent observational results constrain
the class of potentials since they require an energy density $\Ompo
=0.7 \pm .1$ with an equation of state parameter $\wpo =p_{\phi o} /
\rho_{\phi o}\leq -0.6$, where the subscript $o$ refers to present
day quantities \ci{SN1a},\ci{CMBR},\ci{w}.

In this work we show that a non-abelian gauge group with $N_c$ the
number of colours and $N_f$ that of chiral fields leads to an
acceptable quintessence potential. We show that the only degrees of
freedom are precisely the simple choice of $N_c$, $N_f$ and the
number of dynamically generated bilinear fields   which set the
scale of condensation and the power in the potential of the scalar
field responsible for present day acceleration of the universe. Of
course, we are not able to determine from first principles the
values of $N_c, N_f$ but they are at the same footing as the choice
of gauge groups and matter content of the standard model.

The model is quite simple, we start with a non-abelian gauge group
at a high energy scale (could be the unification scale of the
standard model gauge groups) with massless matter fields and we
let it evolve to lower scales. By lowering the energy scale, the
gauge coupling constant becomes large and all fields become
strongly interacting at the condensation scale $\Lm_c$. Below this
scale there are no more free elementary fields, chiral nor gauge
fields, similar to what happens with QCD and we are left with
gauge singlets bilinear fields $\phi^2\equiv <Q\tilde Q >$ (the
square in $\phi$ is to give the field a mass dimension one). We
use Afflek-Seiberg superpotential \ci{Affleck} to determine the
form of the scalar potential $V$ in terms of $\phi$ (related work
can be seen in \ci{Binetruy}). Afterwards, we solve Einstein's
General Relativity equations in a Friedmann--Robertson--Walker
flat metric and determine the cosmological evolution of $\phi$. We
show that a positive accelerating universe at present time with
$\Ompo \simeq 0.7$ and $\wpo < -0.6$ is possible. We will bear in
mind that the second restriction can be set in terms of an
effective equation of state parameter $w_{eff}\equiv\int da\
\Om(a)\wp(a)/\int da\ \Om (a) < -0.7$ \ci{w}.

Furthermore, we constrain the model to have the same unification
scale and gauge coupling as the standard model gauge groups. This is
by all means not a necessary condition but it gives a very
interesting model. We could think of this model as coming from
string theory after compactifying the extra dimensions. The gauge
coupling is unified for all gauge groups, the standard and  non
standard model gauge groups, at the compactification scale which is,
in this case, also the unification scale. We then allow all fields
to evolve cosmologically. Since at the beginning all fields are
massless  they behave as radiation until  a gauge group becomes
strongly coupled and there is a phase transition. Below this scale
the particles charged under the strongly coupled gauge group
condense while the other fields still evolve as radiation. Finally,
we take into account the matter domination period and determine
today's relevant cosmological quantities.

 Let us begin by writing the scalar potential for a non-abelian
$SU(N_c)$ gauge group with $N_f$ (chiral + antichiral) massless
matter fields. The superpotential is given by \newline
$W=(N_c-N_f)(\fr{\Lm^b}{det <Q\tilde Q>})^{1/(N_c-N_f)} $
\ci{Affleck} and the scalar potential in global supersymmetry is
$V=|W_\phi|^2$, with $W_\phi=\pp W/\pp \phi$, giving
\be
V=(2 N_f)^2\Lm_c^{4+n}\phi^{-n}
 \la{v} \ee
where we have taken $det <Q\tilde Q>=\Pi_{j=1}^{N_f} \phi^2_j$,
$n=2\fr{N_c+N_f}{N_c-N_f}$ and $\Lm_c$ is the condensation scale
of the gauge group $SU(N_c)$. We have taken $\phi$ canonically
normalized, however the full Kahler potential $K$ is not known and
for $\phi \simeq 1$ other terms may become relevant \ci{Binetruy}
and could spoil the runaway and quintessence  behaviour of $\phi$.
Expanding the Kahler potential as a series power
$K=|\phi|^2+\Sigma_{i}a_i |\phi|^{2i}/2i$ the canonically
normalized field $\phi'$ can be approximated\footnote{The
canonically normalized field $\phi'$ is defined as
$\phi'=g(\phi,\bar{\phi})\phi$ with $K_\phi^\phi=(g+\phi g
_\phi+\bar{\phi}g_{\bar{\phi}})^2$} by
$\phi'=(K_\phi^\phi)^{1/2}\phi$ and  eq.(\ref{v}) would be given
by $V=(K_\phi^\phi)^{-1}|W_\phi|^2=(2 N_f)^2\Lm_c^{4+n}\phi^{-n}
(K_\phi^\phi)^{(n/2-1)}$. For $n<2$ the exponent term of
$K_\phi^\phi$ is negative so it would not spoil the runaway
behaviour of $\phi$.

In terms of the evolution of the gauge coupling constant we have
 \be
 \Lm_c= \Lm_0 e^{-\fr{1}{2 b_0 g^2_0}}
\la{lm} \ee
 with
$\Lm_0, g_0$ the energy scale and coupling constant at a high
energy scale where the gauge group is weakly coupled and
$b_0=(3N_c-N_f)/16\pi^2$ the one-loop beta function. We would like
to take $\Lm_0$ as the unification scale $\Lm_{gut}\simeq 10^{16}
GeV$ and $g_0$  as the unification coupling
$g_{gut}=\sqrt{4\pi/25.7}$ \ci{unif}.

The presence of the field $\phi$ with potential eq.(\ref{v})
begins only at the  condensation scale $\Lm_c$. We can relate the
scale $\Lm_c$ to the Hubble constant using
$H^2=\rho/3\simeq\Lm_c^4/3$ giving $\Lm_c\simeq (3H^2)^{1/4}$
where we have set the reduced Planck mass to one (i.e. $m_p^2=8\pi
G=1$). By dimensional analysis we set the initial condition for
$\phi$ to be $\phi_i=\Lm_c$ which is the natural choice.

The cosmological evolution of inverse power potential has been
studied in \ci{tracker},\ci{1/q}. The equations to be solved, for a
spatially flat Friedmann--Robertson--Walker  Universe, in the
presence of a barotropic fluid,   which can be either radiation or
matter given by an energy density $\rho_\gm$, are given by
\ci{liddle},\ci{mio.scalar}
 \bea x_N&=& -3 x +
\sqrt {3 \over 2} \lambda\,  y^2 + {3 \over 2} x [2x^2 + \gm_{\gm}
(1 - x^2 - y^2)] \non \\ y_N&=& - \sqrt {3 \over 2} \lambda \, x\, y
+ {3 \over 2} y [2x^2 + \gm_{\gm} (1 - x^2 - y^2)]
 \la{cosmo1} \\
H_N&=& -{3 \over 2} H [2x^2 + \gm_{\gm} (1 - x^2 - y^2)]
 \non \eea
where $N$ is the logarithm of the scale factor $a$, $N \equiv ln
(a)$, $f_N\equiv df/dN$ for $f=x,y,H$ and  $\gm_{\gm}=4/3, 1$ for
radiation or matter fields respectively and $\lambda (N) \equiv -
V' / V$. We have defined the variables $x \equiv \dot \phi / \sqrt
6 H$, $y \equiv \sqrt V / \sqrt 3 H$. In terms of $x, y$ one has
$\Omp =x^2+y^2$ and the equation of state for quintessence field is
given by $w_{\phi}=(x^2-y^2)/(x^2+y^2)$. Generic solutions to
eqs.(\ref{cosmo1}) can be found in \ci{mio.scalar},\ci{generic}.

Notice that all model dependence in eqs.(\ref{cosmo1}) is through
the quantity $\lm (N)$. Using the potential given in eq.(\ref{v}) we
have $\lm=\fr{n}{\phi}=n(H_i y_i)^{-1/2}(H y)^{2/n}/3^{1/4}$ where
$i$ stands for initial conditions. Since $\Lm_c \ll m_p$ and
$\phi_i=\Lm_c$ the initial value of $\lm$ is very large $\lm_i =n\,
m_p/\phi_i \gg 1$ and this has interesting consequences.

From eqs.(\ref{cosmo1}), the evolution of $\Omp $ is to drop
rapidly, $\Omp \ll 1$, in about 3 e-folds, i.e. $\delta N \simeq
3$, regardless of its initial value and it remains very small for
a large period of time (see fig.(\ref{fig2})).  These properties
are due to the fact that $\lm_i$ is large. The evolution of $\phi$
enters a scaling regime with $\lm =constant$ and $\Omp \ll 1$
during all this period. The scaling regime ends when $x=O(1/10)$
and $\Omp $ becomes also of the order of 0.1.

\begin{figure}[ht!]
\begin{center}
\includegraphics[width=8cm]{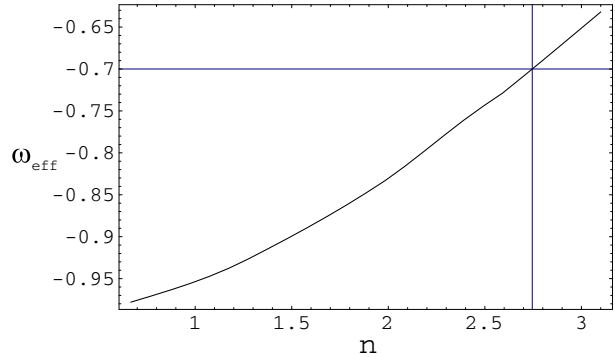}
\end{center}
\caption{\small{Restriction on $n$ from the upper limit $\weff \leq
-0.7$.}} \label{fig1}
\end{figure}

\begin{figure}[ht!]
\begin{center}
\includegraphics[width=8cm]{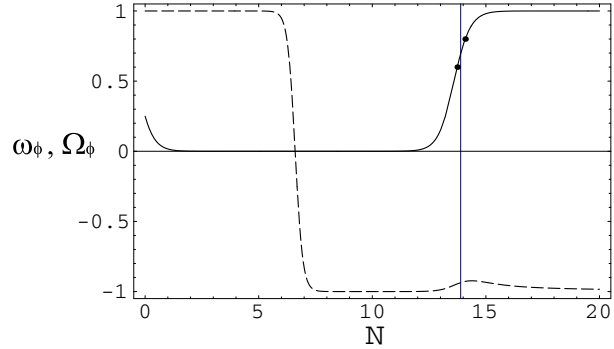}
\end{center}
\caption{\small{Evolution of $\omega_\phi$ and $\Omega_\phi$
(dotted and solid lines respectively). The vertical line represents
the point for which $\Omega_\phi=\Omega_{\phi o}=0.7$ and
$h=h_{0}=0.7$. The lower dot marks $\Omega_\phi=0.6$ while the
upper one stands for $\Omega_\phi=0.8$.}} \label{fig2}
\end{figure}

In this work we have determined $\weff$ for different values of $n$
and we have concluded  that for $\Om_{\phi i}\leq 0.25$ one needs $n
< 2.7$ for $\weff$ to be smaller than $-0.7$ as required \ci{w}. In
Figure \ref{fig1} we show $n$ as a function of $\weff$ assuming
$\Ompo =0.7$ and $h_o=0.7$, where the Hubble constant is given by
$H=100\,h\,km/Mpc \,sec$ . This result constrains many inverse power
models. In fact for $N_c > N_f$ one has $n > 2$. It is important to
point out that we find a decreasing value of $\wpo$ with decreasing
value of $n$ in contrast with \ci{tracker}. The main difference may
be that in our models the value of $\Omp =0.7$ is reached before
$\wp$ joins the tracker solution.

Furthermore, to avoid any conflict with the standard big bang
nucleosynthesis  (NS) results one requires $\Omp (NS) < 0.1$ at the
energy scale of NS \ci{NS}, i.e. $E_{NS}=0.1-10 MeV$. The condition
of not spoiling the NS results rules out the values of $n$ between
$1.2<n<2.1$ for models with $\Om_{\phi i} > 0.1$. This is because
for those values of $n$ the initial value $H_i$ lies within the
value of the Hubble parameter at NS and $\Omp $ is not yet smaller
than 0.1. Of course we could start with a small value of $\Omp $ but
we would loose our democratic choice of initial conditions. These
constraints would leave a window as small as $2.1 < n < 2.7$ for
$N_c>N_f$. However, if we insist deriving $\Lm_c$ from eq.(\ref{lm})
with $\Lm_0$, $g_0$ the unification scale and coupling, the above
conditions constrains the models even more. In fact for $N_c > N_f$
($n>2$) there are no models available that satisfy all constraints:
$\Ompo =0.7$, $\weff < -0.7$ and $\Lm_c$ given by eq.(\ref{lm}) with
$\Lm_0=\Lm_{gut}$ and $g_0=g_{gut}$. A full analysis of all cases
will be presented elsewhere \ci{chris2}.

In order to have  $\Lm_0=\Lm_{gut}$ and $g_0=g_{gut}$  we require
the number of dynamically bilinear fields $Q\tilde Q$ to be
different from $N_f$. Some of these fields may be fixed at their
condensate constant vacuum expectation value (v.e.v.) with $<Q\tilde
Q>=\Lm_c^2$ or we could have a gauge group with unmatching number of
chiral and antichiral fields.

Here, we will present the case of an $SU(3)$ gauge group with
$N_f=6$ chiral fields in the chiral and antichiral representation
and we will assume that only one bilinear field $\phi^2=Q\tilde Q$
becomes dynamical with all other condensates remaining constant with
a v.e.v. equal to $\Lm_c$. Notice that this gauge group is self dual
($\tilde N_c=N_f-N_c=3$ with  $N_f$ flavours)  under Seiberg's
duality transformation\footnote{It is important to point out that
even though it has been argued that for $N_f>N_c$ there is no
non-perturbative superpotential $W$ generated \ci{Affleck} this is
not always the case \ci{ax.asy}.} \ci{duality}.

The potential generated in this case is
\be
V=4 \Lm^{4+n} \phi^{-n}
 \la{v2/3} \ee
with $n=2(1+\fr{2}{N_c-N_f})= 2/3$ and $\phi_i=\Lm_c$. Using
eq.(\ref{lm}) with $16\pi^2 b_0=3N_c-N_f=3$ one has $\Lm_c=4\times
10^{-8} GeV$, which is well below NS energy scale. Notice that
$n<2$ so the noncanonically terms in $K$ will not spoil the
quintessence behaviour of $\phi$ and the mass is $m\simeq H_o$ so
it is cosmologically fine \ci{carol}.

Solving eqs.(\ref{cosmo1}) with the potential given in
eq.(\ref{v2/3}) and initial conditions $\Om_i=0.25$ and
$H_i=(4\Lm^4_c/3 y_i^2)^{1/2}=1 \times 10^{-33} GeV $ gives for
$h_o=0.7$ the values $\Ompo =0.7$ and an equation of state parameter
$\wpo = -0.97$ (with an effective $w_{eff}= -0.98$). We see that the
present day value of the parameters agrees with the analysis of
recent observations \ci{w} and there is no conflict with
nucleosynthesis, since during nucleosynthesis the $SU(3)$ gauge
group was not strongly coupled and all those fields were massless
and behaved as radiation at that epoch. The choice of initial
conditions is not very sensitive and we took it as $\Om_{\phi
i}=0.25$ to be democratic with the standard model gauge groups. A
variation of $40\%$ in the initial value of $\Om_{\phi i}$ gives
still a final result within the range of $h_o=0.7\pm 0.1$ and
$\Ompo=0.7\pm 0.1$. Finally, we show in Figure \ref{fig2} the
evolution of $\Omp$ and $\wp$ as a function of $N$.

To conclude, we have shown that starting with a non-abelian gauge
group with a gauge coupling constant unified with the standard
model gauge couplings  at the unification scale, a gauge singlet
bilinear field $\phi$, arising  due to non-perturbative effects of
to the strongly interacting non-abelian gauge group at the
condensation scale,  gives an acceptable phenomenology for the
cosmological constant and it is therefore a natural candidate for
quintessence.

This work was supported in part by CONACYT project 32415-E and
DGAPA, UNAM project IN-110200.

\thebibliography{}
\footnotesize{

\bib{SN1a} {A.G. Riess {\it et al.}, Astron. J. 116 (1998) 1009; S.
Perlmutter {\it et al}, ApJ 517 (1999) 565; P.M. Garnavich {\it et
al}, Ap.J 509 (1998) 74.}

\bib{CMBR} {P. de Bernardis {\it et al}. Nature, (London) 404, (2000)
955, S. Hannany {\it et al}.,Astrophys.J.545 (2000) L1-L4}

\bib{tracker} I. Zlatev, L. Wang and P.J. Steinhardt, Phys. Rev.
Lett.82 (1999) 8960;  Phys. Rev. D59 (1999)123504

\bib{w}{S. Perlmutter, M. Turner and  M. J. White,
Phys.Rev.Lett.83:670-673, 1999; T. Saini, S. Raychaudhury, V. Sahni
and  A.A. Starobinsky, Phys.Rev.Lett.85:1162-1165,2000 }

\bib{Affleck}{I. Affleck, M. Dine and N. Seiberg, Nucl. Phys.B256
(1985) 557}

\bib{Binetruy}{P. Binetruy, Phys.Rev. D60 (1999) 063502, Int. J.Theor.
Phys.39 (2000) 1859; A. Masiero, M. Pietroni and F. Rosati, Phys.
Rev. D61 (2000) 023509}

\bib{ax.asy}{C.P. Burgess, A. de la Macorra, I. Maksymyk and F. Quevedo
Phys.Lett.B410 (1997) 181}

\bib{unif}{U. Amaldi, W. de Boer and H. Furstenau, Phys. Lett.B260
(1991) 447, P.Langacker and M. Luo, Phys. Rev.D44 (1991) 817}

\bib{1/q} {P.J.E. Peebles and B. Ratra, ApJ 325 (1988) L17; Phys. Rev. D37 (1988) 3406}

\bib{liddle}E.J. Copeland, A. Liddle and D. Wands, Phys. Rev. D57 (1998) 4686

\bib{mio.scalar}{A. de la Macorra and G. Piccinelli, Phys.
Rev.D61 (2000) 123503}

\bib{generic}  A.R. Liddle and R.J. Scherrer, Phys.Rev.
D59,  (1999)023509

\bib{carol} {S.M. Carroll, Phys. Rev.Lett.81 (1998) 3067}

\bib{ax.coinci}{A. de la Macorra, Int.J.Mod.Phys.D9 (2000) 661 }

\bib{NS} {K. Freese, F.C. Adams, J.A. Frieman and E. Mottola, Nucl. Phys. B 287
(1987) 797; M. Birkel and S. Sarkar, Astropart. Phys. 6 (1997)
197.}

\bib{chris2}{A. de la Macorra and C. Stephan-Otto, in preparation}

\bib{duality}{K. Intriligator and  N. Seiberg, Nucl.Phys.Proc.Suppl.45BC:1-28,1996

}
\end{document}